\documentclass{ws-p9-75x6-50}

\begin{document}
\def\be{\begin{eqnarray}}
\def\ee{\end{eqnarray}}
\newcommand{\mat}{\left ( \begin{array}{cc}}
\newcommand{\emat}{\end{array} \right )}
\title{Effective Low Energy Theories and QCD Dirac Spectra}
\author{D. Toublan and J.J.M. Verbaarschot}

\address{Department of Physics and Astronomy, SUNY Stony Brook, \\
Stony Brook, NY 11790, USA,\\
e-mail: toublan@nuclear.physics.sunysb.edu,\\ 
e-mail: verbaarschot@nuclear.physics.sunysb.edu}

\maketitle

\abstracts{We analyze the smallest Dirac eigenvalues by formulating
an effective theory for the QCD Dirac spectrum. We find that in
a domain where the kinetic term of the effective theory can be ignored,
the Dirac eigenvalues are distributed according to a Random Matrix Theory
with the global symmetries of the QCD partition function. The kinetic
term provides information on the slope of the average spectral density
of the Dirac operator.
In the second half of this lecture we interpret quenched
QCD Dirac spectra at nonzero chemical potential  
(with eigenvalues scattered in the complex plane)
in terms of an effective low energy theory.
}

\section{Introduction}
There are two essentially different approaches to many-body problems. 
The first approach is to study them by means of Monte-Carlo
simulations of the microscopic theory. The second approach is to
isolate the relevant degrees of freedom and to describe them by
an effective theory. Both approaches have their merits and generally
their complementarity leads to a deeper understanding of the underlying
phenomena. In this lecture we will focus on 
Quantum Chromo-Dynamics (QCD) which is the theory of the strong 
interactions. A great deal of effort has been devoted to Monte
Carlo simulations of lattice QCD~\cite{DeTar}. They provide a firm footing
for our understanding of
nonperturbative phenomena such as confinement and chiral symmetry breaking.
For QCD at low energy an alternative approach is possible.
Because of spontaneous breaking
of chiral symmetry, QCD at low energy reduces to a theory of weakly
interacting Goldstone bosons. Although this theory cannot be derived from
QCD by means of an ab-initio calculation, its Lagrangian 
is determined uniquely by chiral symmetry and Lorentz 
invariance~\cite{ChPTfound}.
The validity of this low-energy theory is based on the presence of a mass-gap
which is a highly nontrivial and nonperturbative feature of QCD.

Chiral symmetry is spontaneously broken at low temperatures. Mainly
through lattice simulations, it is now widely accepted 
that a chiral restoration
transition takes place at a temperature of about 140 $MeV$. The order
parameter of this transition is the density per unit volume 
of Dirac eigenvalues
near zero. This is the reason why we are interested
in the properties of the smallest 
eigenvalues of the Dirac operator~\cite{LS,SVR}.
The situation at finite baryon density and zero temperature is much
less well understood. Monte-Carlo simulations are only possible if the 
fermion determinant is ignored. It has been shown that the so called
quenched approximation  fails spectacularly at nonzero chemical 
potential~\cite{everybody,Misha}. On general grounds it 
is expected that a chiral phase transition occurs at a value of the 
chemical potential where it
becomes advantageous to create the lightest particles with nonzero baryon
number. In QCD these are the nucleons, but in the quenched
approximation, they are Goldstone modes made from quarks and conjugate
anti-quarks~\cite{Misha}. At nonzero chemical potential, 
the relation between the chiral order parameter
and the QCD Dirac spectrum, which is now scattered in the complex
plane, is much less transparent. The failure of quenching can 
be understood as the absence of spectral
ergodicity; at finite density the ensemble averaged Dirac spectrum is 
completely different form the spectral averaged Dirac spectrum. 
We will investigate the Dirac spectrum by means of an effective theory.
Our approach is similar to the one used for QCD with two colors
and for QCD with adjoint quarks~\cite{KST,KSTVZ}.
By investigating the properties of the smallest Dirac eigenvalues
we hope to obtain a better understanding of this problem 
and other related issues.

In this lecture we wish to 
discuss to what extent spectra of the QCD Dirac operator
can be derived from an effective low energy partition function. 
In the first half of this talk we formulate an effective theory for 
the Dirac spectrum based on a spectrum generating function that in
addition to the usual quarks contains bosonic 
ghost quarks to properly normalize
the spectral density. This trick has been widely used in the super-symmetric
formulation of Random Matrix Theory~\cite{Efetov,Brezin,VWZ}. 
We identify a domain where
the results for the resolvent are given by a 
chiral Random Matrix Theory with the symmetries of the QCD partition
function. We find that our results are in agreement with recent 
lattice simulations.
In the second half of this lecture we discuss quenched QCD Dirac
spectra at nonzero chemical potential in terms of  an effective theory.
This lecture is based on several recent 
articles~\cite{Vplb,Osbornprl,Osbornnpb,OTV,DOTV,TV} and additional
background material can be found in several recent 
reviews~\cite{HDgang,Montambaux,kyoto}. 

\section{The QCD partition function}
The QCD partition function for $N_f$ quarks with mass $m$, temperature $T$ and
quark chemical potential $\mu$ is given by
\be
Z(m,\mu,T) ={\rm Tr\,} e^{-\frac{H_{\rm QCD} -\mu N}T},
\ee
where $H_{\rm QCD}$ is the Hamiltonian of QCD and $N$ is the quark number
operator. The trace is over all states of the theory. This partition 
function can be rewritten as a Euclidean Feynman path integral
\be
Z(m,\mu,T) = \langle {\det}^{N_f} (D+m) \rangle_{\rm YM},
\ee
where the average of the fermion determinant is over the 
Euclidean Yang-Mills action and
the Dirac operator is denoted by $D$. In this lecture we focus
on the basic symmetry properties of $D$ and its
explicit representation is not necessary. 
For simplicity we take all $N_f$ quark masses equal to $m$.

\subsection{The eigenvalues of the Dirac operator}
The main subject of this talk are the properties of the smallest eigenvalues
of the QCD Dirac operator. In a chiral representation of the gamma-matrices,
the Euclidean Dirac operator has the block structure
\be
D = \mat 0 & id + \mu \\ id^\dagger +\mu & 0 \emat,
\ee
where $d$ is the covariant derivative of the color group.
For $\mu \ne 0$ the Dirac operator does not have any hermiticity properties
and the eigenvalues will be scattered in the complex plane. We will return
to this case in the second half of this talk. For $\mu =0$ the Dirac operator
is anti-Hermitian, $D^\dagger = -D$ and has purely imaginary eigenvalues.
The nonzero eigenvalues occur in pairs $\pm i\lambda_k$, whereas zero 
eigenvalues related to the topology of the gauge fields remain unpaired.
For simplicity we will restrict the discussion in this lecture to gauge
fields with a trivial topology where all  eigenvalues occur as pairs.
For example, our gauge fields could be composed of fields of an
equal number of instantons and anti-instantons~\cite{SVR}.
In terms of its eigenvalues, the Euclidean partition function can be written
as
\be
Z(m,\mu,T) = \left \langle \prod_k(i\lambda_k + m)^{N_f} \right 
\rangle_{\rm YM}.
\ee

The order parameter of the chiral phase transition is the chiral 
condensate
\be
\Sigma &=& 
               \lim_{m\rightarrow 0} \lim_{V\rightarrow \infty}
                         \frac 1{VN_f} \partial_m \log Z\nonumber\\ 
                       &=&\lim_{m\rightarrow 0} \lim_{V\rightarrow \infty}
\frac 1V \sum_{k} 
\left \langle \frac 1{m+i\lambda_k}\right \rangle_{\rm QCD} 
                       = \lim_{m\rightarrow 0} \lim_{V\rightarrow \infty}
\frac 1V \sum_{\lambda_k > 0} \left \langle \frac {2m}{m^2+\lambda_k^2}
                      \right  \rangle_{\rm QCD},
\ee
where the average with label QCD includes both the fermion determinant
and the Yang-Mills action. 
At finite volume, the chiral condensate is zero for $m=0$. Only
if the thermodynamic limit is taken before the chiral limit ($m\rightarrow 0$)
can the chiral condensate become nonzero. If this is the case, chiral
symmetry is broken spontaneously. The situation is analogous to the
magnetization in a Heisenberg model. At finite volume and zero magnetic
field the magnetization is zero. Spontaneous magnetization arises if the
thermodynamic limit is taken before putting the external field to zero.
According to Goldstone's
theorem spontaneous breaking of a continuous symmetry 
implies the existence of massless Goldstone bosons.
Because of confinement QCD has a mass gap and, at low energy, the 
partition function is dominated by the Goldstone modes. The effective
theory describing their interactions then
follows from chiral symmetry and Lorentz invariance. 

One of the questions we would like to address in this lecture is
whether there is a relation between the existence of a well-defined
low-energy theory and the spectrum of the Dirac operator. A second,
seemingly unrelated question is the connection of this low-energy
with Random Matrix Theory.
 
\section{Symmetries of the Spectrum Generating Function at $\mu=0$}
We will study the Dirac spectrum by means of the resolvent defined by
\be
\Sigma(z) = \frac 1V \left \langle {\rm Tr\,} \frac 1{D+z} \right
\rangle_{\rm QCD}.
\ee
The mass $z$ in the resolvent is not related
to the mass $m$ in the fermion determinant. Such mass is known in the
literature as a valence quark mass. However, we will use the more
appropriate name of spectral mass. 

In terms of the spectral density 
\be
\rho(\lambda) = \langle \sum_k \delta(\lambda -\lambda_k)\rangle_{\rm QCD},
\ee
the resolvent can be written as
\be
\Sigma(z) = \frac 1V \int \rho(\lambda) d\lambda \frac 1{i\lambda +z}.
\ee
Therefore, $\Sigma(z)$ is an analytic function with a cut on the 
imaginary axis. 
This identity can be inverted by taking the discontinuity across the cut
\be
\frac{\rho(\lambda)}V =\lim_{\epsilon \rightarrow 0}
\frac 1{2\pi} (\Sigma(i\lambda +\epsilon) - \Sigma(i\lambda -\epsilon)).
\ee

In order to obtain a generating function for $\Sigma(z)$ we use a method
that has been widely used in the theory of 
disordered systems~\cite{Efetov,VWZ}, namely~\cite{pqChPT,OTV,DOTV}
\be
Z^{\rm sp }(z,J,m)  ~=~ 
\langle \frac{\det(D +z+J)}{\det(D +z)} {\det}^{N_f}(D + m) \rangle_{\rm YM}.
\label{pqQCDpf}
\ee
The resolvent is then given by
\be
\Sigma(z) = \left . \frac1V \partial_J Z(z,J,m)\right |_{J=0}.
\ee 
The spectrum generating function (\ref{pqQCDpf}) contains $N_f+1$ fermionic
quarks and one bosonic quark. In addition to the chiral symmetry, this 
partition function also contains a super-symmetry that mixes fermionic and
bosonic quarks. This can be seen by
rearranging the fermion determinant as
\be
\det \mat m & id \\ id^\dagger & m \emat = 
\det \mat  id  & m \\ m &id^\dagger  \emat. 
\ee
We observe  that for $m=z=J=0$ the partition function is invariant under
\be
\left (\begin{array}{cccccc} id &&&&& \\
                              & \ddots &&&&\\
                              && id &&& \\
                               &&& id^\dagger &&\\
                               &&&& \ddots & \\
                               &&&&& id^\dagger 
\end{array} \right ) 
\rightarrow \mat U & \\ & V \emat 
\left (\begin{array}{cccccc} id &&&&& \\
                              & \ddots &&&&\\
                              && id &&& \\
                               &&& id^\dagger &&\\
                               &&&& \ddots & \\
                               &&&&& id^\dagger 
\end{array} \right ) 
\mat U^{-1} & \\ & V^{-1} \emat,
\ee
where  $U$ an $V$ are $(N_f+1|1) \times (N_f+1|1)$ super-matrices.
Mathematically, this symmetry group is known as 
$Gl(N_f+1|1) \times Gl(N_f +1 | 1)$.
A $Gl(1)$ subgroup is broken by the anomaly. The chiral condensate
$\Sigma \equiv \langle \bar \psi \psi \rangle$ 
is only invariant under the diagonal
subgroup  with $U=V$. Therefore the symmetry is broken spontaneously
to $Gl(N_f+1|1)$. As is the case in QCD, the masses of the Goldstone
modes are given by the Gell-Mann-Oakes-Renner relation
\be
\frac {\sqrt {2m\Sigma}}{F}, \qquad \frac {\sqrt{(m+z)\Sigma}}{F}, \qquad
\frac {\sqrt{2z\Sigma}}{F},
\label{pions}
\ee
where $F$ is the pion decay constant.
The Goldstone modes corresponding to a fermionic and a bosonic quark are
fermionic whereas all other Goldstone modes are bosonic. 

\section{Effective Low-Energy Theory}
The manifold $Gl(N_f+1|1)$ is not Riemannian, and is therefore not suitable
as a Goldstone manifold. The Goldstone manifold is given by the maximum
Riemannian submanifold of the symmetric superspace $Gl(N_f+1|1)$
which will be denoted by $\hat {Gl}(N_f+1|1)$~\cite{Zirnall,OTV,DOTV}. 
If we ignore
certain complications related to the topological structure of the 
QCD vacuum, this
partition function is given by
\be
Z(m,J,z) = \int_{\hat {Gl}(N+f+1|1)}  dU e^{- \int d^4 x
\left [\frac{F^2}4{\rm Str\, }\partial_\mu U \partial_\mu U^{-1} -
\frac 12 \Sigma {\rm Str\,} M (U+U^{-1}) \right ]},
\ee
where the mass matrix $M= {\rm diag} (m, \cdots, m, z+J, z)$.The super-matrix $U$ is
parameterized as~\cite{OTV,DOTV} 
\be
U = \mat V & \alpha \\ \beta & e^s \emat \equiv e^{i\sqrt 2 \Phi/F}.
\ee
Here, $V$ is a $U(N_f+1)$-matrix, $\alpha$ and $\beta$ are Grassmann valued
vectors of length $N_f+1$ and $s$ is a real number.

In order to estimate the relative importance of the two terms in the
effective Lagrangian we expand the fields
to second order in the pion fields $\Phi =\pi^a t_a$,
\be
{\cal L}^{\rm eff} = \frac 12 \partial_\mu \pi^a \partial_\mu \pi^a 
+\frac 12 M_a^2 \pi_a^2,
\ee
where $M_a$ is one of the Goldstone masses given in (\ref{pions}). In a box
of volume $L^4$, the 
smallest nonzero momenta are of the order $p_\mu \sim 1/L$. 
Let us consider QCD in the chiral limit (with $m =0$). Then the regular
mesons are massless whereas mesons containing one or two spectral quarks
have a nonzero mass given by (\ref{pions}).   
Therefore, if~\cite{Vplb}
\be
\frac {z\Sigma}{F^2} \ll \frac 1{L^2} 
\ee
the correlation functions involving spectral quarks are
dominated by contributions from the zero momentum Goldstone modes. 
Nonperturbatively, the partition function reduces
to a group integral in this limit. This integral has been calculated
analytically~\cite{OTV,DOTV} resulting in the following dependence of the
condensate on the spectral mass
\be
\frac{\Sigma(z)}{\Sigma}=  {\mu_z} \left [ I_{N_f}(\mu_z) K_{N_f}(\mu_z)
+ I_{N_f+1}(\mu_z) K_{N_f-1}(\mu_z) \right ] ~.
\label{valcond}
\ee
As already could be observed from the generating function, 
it depends on $z$ only
through the combination
$\mu_z \equiv z V \Sigma$.  This result was first obtained~\cite{Vplb} 
by means of chiral Random Matrix Theory to be discussed
in the next section.

QCD is not the only theory that reduces to this effective partition function.
In fact, any partition function with the same pattern of chiral symmetry 
breaking and a mass gap reduces to the same low-energy theory.
Explicit examples are given by the random flux model~\cite{AS} and
other models with a hopping term, disorder and chiral 
symmetry~\cite{Takahashi,GWW}. 
A natural question is what is the simplest theory with the
same zero momentum sector as QCD. This theory is chiral Random Matrix Theory.

\section{Chiral Random Matrix Theory}

In the sector of topological charge $\nu$ and  for $N_f$ quarks with mass $m$, 
the chiral random matrix partition function
 with the global symmetries of the QCD partition function
is defined by~{\cite{SVR,V}}
\be
Z^\nu_\beta({\cal M}) =
\int DW \prod_{f= 1}^{N_f} \det{\left (\begin{array}{cc} m & iW\\
iW^\dagger & m \end{array} \right )}
e^{-\frac{N \beta}4 \Sigma{\rm Tr\,}W^\dagger W },
\label{ranpart}
\ee
where $W$ is a $n\times (n+|\nu|)$ matrix and $N= 2n+|\nu|$.
As is the case in QCD, we assume that the equivalent of the topological charge
$\nu$ does not exceed $\sqrt N$,
so that, to a good approximation, $n = N/2$.
Then the parameter $\Sigma$ can be identified as the chiral condensate and
$N$ as the dimensionless volume of space time (Our units are defined
such that the density of the modes $N/V=1$).
The matrix elements of $W$ are either real ($\beta = 1$, chiral
Gaussian Orthogonal Ensemble (chGOE)), complex
($\beta = 2$, chiral Gaussian Unitary Ensemble (chGUE)), or quaternion real
($\beta = 4$, chiral Gaussian Symplectic Ensemble (chGSE)). 
For QCD with three or more colors and quarks in the fundamental representation
the matrix elements of the Dirac operator are complex and we have $\beta = 2$.
The ensembles with $\beta =1 $ and $\beta =4$ are relevant in the case
of two colors and adjoint fermions, respectively.
The reason for choosing a Gaussian distribution of the matrix elements is
its mathematical simplicity. It can be shown that the correlations
of the eigenvalues on the scale of the level spacing do not depend on
the details of the probability distribution~
\cite{brezin,ADMN,Sener1,GWu,Seneru,Kanzieper,Dampart,Damtop,Damint,mag}. 

Together with the Wigner-Dyson ensembles and
the superconducting random matrix ensembles~\cite{Altland}
the chiral ensembles can be classified according to the Cartan classification
of large symmetric spaces~\cite{Zirnall}.
 
\section{Scales in the Dirac Spectrum}

For a nonzero value of the chiral condensate $\Sigma$ we can identify 
three important scales in the Dirac spectrum. The 
first scale is the smallest nonzero eigenvalue of the Dirac operator
given by $\lambda_{\rm min} =1/\rho(0)=\pi/\Sigma V$.
The second scale is the spectral mass for which the Compton wavelength of 
the associated Goldstone bosons is 
equal to the size of the box. As discussed above this
scale is given by~\cite{GL,Altshuler,Vplb,Osbornprl} 
\be
E_c = \frac {F^2}{\Sigma L^2}.
\label{Thouless}
\ee
In mesoscopic physics~\cite{Altshuler}
this scale is known as the Thouless energy. It is given by the inverse
diffusion time of an electron through a sample of length $L$.  
In Euclidean QCD, this scale can be interpreted in terms of diffusive motion
of quark in 4 Euclidean dimensions and one artificial time 
dimension~\cite{Osbornprl}.
A third scale is given by a typical hadronic mass scale. The three scales 
are ordered as $\lambda_{\rm min} \ll E_c \ll \Lambda$.

For spectral masses $z \ll E_c$ the kinetic term in the effective 
action can be neglected and the low-energy partition function can be 
reduced to a zero dimensional integral. As already mentioned, 
the results for $\Sigma(z)$ 
obtained from the spectral partition function and from chRMT 
coincide~\cite{OTV,DOTV} in this domain. 

\begin{center}
\begin{figure}[!ht]
\centering\includegraphics[width=55mm,angle=270]
{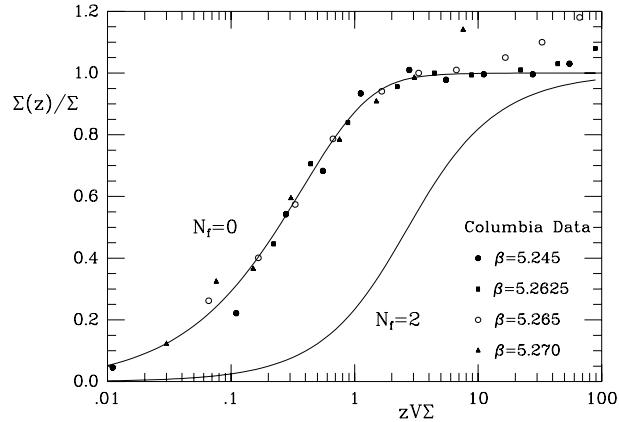}
\caption[]{The spectral mass dependence of the chiral condensate
$\Sigma(z)$ plotted as $\Sigma(z)/\Sigma$ versus $z V\Sigma$.
The dots and squares represent lattice results by the Columbia group
\cite{Christ} for values of $\beta$ as indicated in the label of the figure.}  
\label{fig5}
\end{figure}
\end{center}
In the domain $E_c \ll z \ll \Lambda$ the kinetic term has to be taken
into account. The slope of the Dirac spectrum at
$\lambda = 0$ can be obtained from a one-loop calculation.
For $N_f$ massless flavors the result is given by~\cite{Smilstern,OTV,TV}
\be
\frac{\rho'(0)}{\rho(0)} =
\frac {(N_f-2)(N_f+\beta)}
{16\pi \beta}\frac{\Sigma_0 }{F^4}.
\ee 
Here, $\beta$ denotes the Dyson index of the Dirac operator defined before.

The domain below $E_c$ has been investigated extensively
by means of lattice QCD 
simulations and complete agreement with the chRMT results has been found~
\cite{Vplb,Tilo,Hip,edwardsprl,Heller,hiplat99,Tiloval,Poulval,Karlval}$^,$
\cite{Lang,topo,BergMPW,Tiloch,Damedhn,Damhnr,Edwards,karllat,FarHip}.
In Fig. 1 we show a comparison of the ratio $\Sigma(z)/\Sigma$ versus 
$zV\Sigma$ of lattice results obtained by the Columbia group~\cite{Christ}
and eq.  (\ref{valcond}) for $N_f =0$ and 2. The point where the
lattice data depart from the chRMT result roughly coincides with the 
scale $E_c$ defined in eq. (\ref{Thouless}).

\section{QCD Dirac spectra at $\mu \ne 0$}

The QCD partition function at $\mu\ne 0$ is given by
\be 
Z = \sum_\alpha e^{-\frac{ E_\alpha - \mu N_\alpha}T}.
\ee
Below we will derive the Dirac spectrum from the property that
for $T\rightarrow 0$ only states with $E_\alpha/N_\alpha < \mu$
contribute to the partition function. 

For $\mu \ne 0$ the eigenvalues are scattered in the complex plane. 
The spectral density is then given by~\cite{Girko}
\be
\rho(z) =\frac 1\pi \partial_{z^*} G(z),
\ee
with the resolvent $G(z)$ defined as usual by
\be
G(z) = \left \langle \frac 1V {\rm Tr\,} \frac 1{D+z} {\det}^{N_f} (D+m) 
\right \rangle_{\rm 
YM}
\ee
Writing out explicitly the real and imaginary parts, we observe that $G(z)$ is
the  
electric field in the plane of charges located at the positions of the 
eigenvalues. The spectrum of the Dirac operator has been studied
extensively in the context of 
Random Matrix Theory~
\cite{Misha,halaszyl,Fein,Nowak,supernonh,fyodorov,Efetovnh,Osbornmu,phase}
$^,$
\cite{glas,efetovsym,fyodorovpoly,benoit}. Below, we will
obtain the quenched Dirac spectrum using an effective low energy.
To lowest order in $\mu^2$ our results agree with quenched Random
Matrix Theory.

At $\mu \ne 0$ a new ingredient enters in the definition of the
the spectrum generating function. In order
to assure convergence of the bosonic integral we need an additional factor
$\det (D^\dagger + z^*)$ in the denominator, and a corresponding factor
in the numerator,
\be
Z(z, m) = \left \langle\frac{\det(D+z+J) \det(D^\dagger +z^*+J^*) 
{\det}^{N_f}(D+m)}
                   {\det(D+z)\det(D^\dagger+z^*)}\right\rangle_{\rm YM}.
\ee
The additional factor can be interpreted in terms of conjugate quarks with
a baryon number opposite to that of ordinary quarks~\cite{Misha}. This 
opens the possibility of baryonic Goldstone modes consisting of a quark
and a conjugate anti-quark. For simplicity we only discuss the quenched case
where $N_f = 0$.

In order to construct the effective partition function we have to identify
the symmetries of the spectrum generating function. To this end we
rewrite the product of the determinant and the conjugate determinant as
follows
\be
\det(D+z) \det(D+z^*) &=& \det \mat id+\mu & z \\ z & id^\dagger+\mu \emat
\det \mat -id+\mu & z^* \\ z^* & -id^\dagger+\mu \emat \nonumber \\
&=& \det \left ( \begin{array}{cccc}
                   id+\mu &0 & z & 0 \\
                   0 & id-\mu & 0 & -z^* \\
                   z & 0 &id^\dagger+\mu &0 \\
                   0 & -z^* &0 &id^\dagger-\mu   \end{array} \right )
\ee
As can be observed from the sign 
of the chemical potential, the baryon  number of the conjugate 
quarks is opposite to that
of the usual quarks. 

In this case the chiral symmetry is broken spontaneously according to
\be
Gl(2|2) \times Gl(2|2) \rightarrow Gl(2|2).
\ee
This symmetry is broken explicitly by the term proportional to $\mu$ 
as
\be
Gl(1,1|1,1) \times Gl(1,1|1,1).
\ee
This gives rise to an additional term in the effective action. In order
to maintain  invariance under both the left and right symmetry groups
we need at least a term of second order in $\mu$ and $U$. One easily verifies
that to this order, the only static 
term with the correct symmetry properties
is given by 
\be
\int d^4 x \mu^2 {\rm Str\,} UT_3 U^{-1} T_3,
\ee
where the symmetry breaking matrix $T_3$ is a diagonal matrix with 
nonzero matrix elements equal to 1 except on the positions corresponding
to the conjugate quarks, where they are -1. For example, for $N_f = 0$ we have
\be
T_3 = \left ( \begin{array}{cccc}
                      1&&&\\ &-1& &\\ &&1& \\ &&& -1
               \end{array} \right ).
\ee
If we ignore complications related to the topology of the gauge field 
configurations, the static part of the effective partition function is given by
\be
\int_{U \in \hat{Gl}(2|2)} e^{ \frac 12 \Sigma V
{\rm Str\,} M(U+U^{-1}) -\frac 12 V \mu^2F^2 {\rm Str\,} UT_3 U^{-1} T_3 },
\ee
where the mass matrix is given by $M={\rm diag}(z+J,z^*+J^*,z,z^*)$.
The integral is over the maximum Riemannian submanifold of $Gl(2|2)$.
The relative coefficient of the two symmetry breaking terms is determined
by the condition that the partition function should have a singularity when
$2\mu$ becomes equal to the mass of the lightest meson. The coefficient of
the term $\sim \mu^2$ has to be chosen such that 
 the effective meson mass vanishes at
\be
\mu^2 = \frac {{\rm Re\,}z \,\Sigma}{2F^2},
\ee
guaranteeing a nonanlyticity of the  $\mu$-dependence at this point.
The value of this coefficient can be obtained more elegantly by
means of a gauge principle. This construction has been performed for
QCD with two colors  and for QCD with adjoint
fermions~\cite{KST,KSTVZ}. 
In addition to the term proportional
to $\mu^2$ we then obtain the coefficients of the terms in the effective 
Lagrangian that are linear in $\mu$ and the momentum. 

Baryonic Goldstone modes contain one ordinary quark and 
one conjugate quark both with  mass $z$ resulting in a square mass
of $2{\rm Re \,}z\, \Sigma/F^2.$
For $\mu^2 < {\rm Re}\,  z \,\Sigma / 2F^2$ only 
the vacuum state contributes to the 
QCD partition function, and thus
\be
Z(z,J)= e^{2{\rm Re \,}J\,\Sigma V}.
\label{cphase}
\ee
In the effective theory this corresponds to the saddle-point $U=1$.
For $\mu^2 > {\rm Re}\,  z\, \Sigma / 2F^2$, baryonic Goldstone modes
contribute to the partition function resulting in a nonzero baryon density.
The density stays finite because of the repulsive interaction between
the Goldstone modes.
In terms of the effective partition function this happens  by the
rotation of the saddle point from $U=1$ to a nontrivial value.
By making the Ansatz for the boson-boson and the fermion-fermion blocks
\be
U_{\rm BB} = U_{FF} = \mat \cos \theta &\sin \theta \\ -\sin \theta & 
\cos \theta \emat,
\ee
one finds that $\cos \theta = {\Sigma {\rm Re}\, z}\,/{2\mu^2 F^2}$.
A similar rotation of the saddle point has been found in the
analysis of nonhermitian random matrix 
models~\cite{fyodorov,supernonh,Efetovnh}. We then find 
the following $J$-dependence of the partition function
\be
Z(z,J) = \exp[\frac{V\Sigma^2 {\rm Re\, }J\,{\rm Re\, }z\,}{\mu^2 F^2}].
\label{dphase}
\ee
The resolvent in both domains follows by differentiation with respect to
the source $J$. For ${\rm Re}\,  z \,> 2\mu^2F^2/\Sigma$ we find from
(\ref{cphase})
\be
G(z) = \Sigma\quad {\rm and} \quad \rho(z) = 0,
\ee
and for ${\rm Re}\,  z \,< 2\mu^2F^2/\Sigma$ the result for the resolvent
following from (\ref{dphase}) is given by
\be
G(z) = \frac{\Sigma^2 {\rm Re}\, z\,}{2\mu^2 F^2} \quad {\rm and} \quad 
\rho(z) = \frac{\Sigma^2 }{4\mu^2 F^2}.
\ee
We observe that the resolvent is continuous at the transition point.
The eigenvalues are thus located inside a strip of width $4F^2 \mu^2/\Sigma$.
In terms of the interpretation of the resolvent as an electric field, an
eigenvalue density that does not 
depend on the imaginary part of $z$ results in a constant electric
field outside of the strip of eigenvalues. 
Ignoring finite size effects
this is indeed what is observed in quenched lattice QCD 
simulations~\cite{everybody}.
In the quenched approximation
these results are in agreement with an explicit Random Matrix
Model calculation. Therefore, to order $\mu^2$ the Random Matrix partition
function reproduces exact QCD results.

\section{Conclusions}
We have shown that the correlations of QCD Dirac eigenvalues 
at a scale well below the hadronic mass scale can be  
obtained analytically. They are given by an effective
theory for the generating function of the resolvent for the QCD Dirac
spectrum. For mass scales below $F^2/\Sigma L^2$ the kinetic term in the 
effective theory can be ignored resulting in eigenvalue correlations
given by chRMT. Our results have been confirmed by
numerous lattice QCD simulations.
The same procedure can be followed for QCD at nonzero chemical potential.
Our results indicate that in the quenched limit the eigenvalues are
scattered in a strip with a width determined by the lightest meson
of our theory. This shows that, to lowest nontrivial 
order in the chemical potential, chiral Random Matrix Theory provides
an exact description of the QCD Dirac spectrum.

\vskip 0.5cm
\noindent {\bf Acknowledgments.} 
We gratefully acknowledge all our collaborators in this
project. A. Altland, B. Simons, M. Stephanov and M. Zirnbauer are
thanked for useful discussions.
This work was partially supported by the US DOE grant
DE-FG-88ER40388. One of us (D.T.) was supported in part by
``Holderbank''-Stiftung and by Janggen-P\"ohn-Stiftung.

\end{document}